\documentclass[manuscript]{jair}

\setcopyright{cc}
\copyrightyear{2025}

\RequirePackage[
  datamodel=acmdatamodel,
  style=acmauthoryear,
  backend=biber,
  giveninits=true,
  uniquename=init
]{biblatex}

\ccsdesc[500]{Computing methodologies~Machine learning}
\ccsdesc[500]{Computing methodologies~Natural language processing}

\usepackage{booktabs}
\usepackage{graphicx}

\makeatletter
\@ifundefined{footnotetextauthorsaddresses}{\newcommand{\footnotetextauthorsaddresses}[1]{}}{}
\makeatother

\begin{document}

\title[Relational Validation of LLMs]{A Multi-Source Framework for Relational Validation of Large Language Models Using Expert-Curated Encyclopedic Sources}

\author{Moses Boudourides}
\email{Moses.Boudourides@northwestern.edu}
\affiliation{%
  \institution{School of Professional Studies, Northwestern University}
  \city{Evanston, IL}
  \country{USA}
}

\begin{abstract}
This paper introduces a novel, multi-source framework for the relational validation of Large Language Models (LLMs). While existing benchmarks have demonstrated LLMs' proficiency at factual recall, their ability to understand and reproduce the intricate web of relationships that defines a domain's conceptual structure remains largely unexplored. Our three-layer analytical framework provides a scalable and robust methodology for assessing the depth of an LLM's knowledge across diverse academic domains. By comparing LLM-generated knowledge graphs to expert-curated encyclopedias, we reveal a consistent and significant ``relational deficit'': LLMs recognize domain-specific concepts but consistently fail to reproduce their relational structure. Our findings highlight the need for more sophisticated evaluation metrics that go beyond simple accuracy and assess the relational integrity of an LLM's knowledge. We demonstrate that this deficit is highly domain-dependent, with performance varying significantly across ten specialized encyclopedias spanning sociology, political science, philosophy, and other fields. The cases of complete relational failure in the most specialized domains are particularly revealing, suggesting that the LLM's internal knowledge representation is not aligned with the conceptual structures of these fields. This has significant implications for the deployment of LLMs in high-stakes applications that require a deep, nuanced understanding of domain-specific knowledge.
\end{abstract}

\keywords{Large Language Models, Relational Validation, Knowledge Graphs, Social Network Analysis, LLM Evaluation}

\maketitle

\section{Introduction}

Large Language Models (LLMs) have demonstrated strong performance across a wide range of natural language processing tasks, including text generation, summarization, and question answering \cite{Zhao2023}. Yet performance on such tasks does not, by itself, establish that a model reliably represents \emph{structured} domain knowledge. In many scholarly domains, understanding is constituted not only by recognizing key concepts but also by reproducing the network of relations that organizes those concepts into an expert-endorsed conceptual system.

Most widely used LLM evaluation benchmarks prioritize factual recall and problem solving. For example, MMLU \cite{Hendrycks2020} and related test suites assess whether a model can answer questions correctly across broad subject areas. These benchmarks are valuable, but they largely probe \emph{local} correctness: isolated facts, short-form explanations, or task-specific outputs. They do not directly measure whether a model preserves the \emph{relational structure} of a domain: which concepts are linked, which concepts serve as hubs, and how cross-references organize meaning.

This paper introduces a scalable framework for \emph{relational validation} of LLMs, using expert-curated encyclopedic sources as a reference standard. We treat each encyclopedia as defining a directed knowledge graph whose nodes are domain-specific terms and whose edges represent curated cross-references. We then prompt an LLM with each term and extract an LLM-induced directed graph from the model's output. Comparing the two graphs yields a multi-layer view of relational fidelity: (i) graph-level structural and relational similarity, (ii) node-level agreement on which concepts are central, and (iii) edge-level recovery of curated links.

Across ten specialized encyclopedias spanning philosophy, political science, history, cultural studies, psychoanalysis, and related areas, we find a consistent \emph{relational deficit}: while the model typically recognizes many individual terms, it fails to reproduce the curated relational structure connecting them. The deficit is strongly domain-dependent. In particular, the most specialized domains exhibit near-zero structural similarity and near-zero link recovery, indicating that the model's induced relational organization is not aligned with expert conceptual structure in these fields.

\section{Literature Review}

Our approach connects three research streams: evaluation of Large Language Models (LLMs), knowledge-graph representations of domain knowledge, and network-analytic comparison of relational structure. This section situates our work within these complementary perspectives and motivates the need for relational validation as a novel evaluation paradigm.

\subsection{Evaluation of Large Language Models}

Recent surveys document both the remarkable capabilities and persistent limitations of contemporary LLMs \cite{Zhao2023,Huang2023}. While these models have demonstrated impressive performance on diverse natural language understanding and generation tasks, concerns about their robustness, hallucination rates, and trustworthy deployment in high-stakes applications remain central to the field. Standard evaluation approaches typically rely on benchmark datasets that measure performance on discrete tasks---question-answering accuracy, semantic similarity, entailment detection, and similar metrics. However, these task-centric benchmarks often fail to capture whether a model has genuinely internalized the relational structure of knowledge within a domain.

The hallucination problem is particularly acute: LLMs can confidently generate plausible-sounding but factually incorrect statements, especially when asked about specialized or domain-specific topics \cite{Zhang2023,Azaria2023}. This suggests that surface-level accuracy on isolated examples does not guarantee that a model has learned the coherent, interconnected web of relationships that characterize genuine domain understanding. Existing benchmarks, while valuable, are typically narrow in scope and may not generalize to the full complexity of expert knowledge in specialized fields.

\subsection{Knowledge Graphs and Structured Knowledge Representation}

Knowledge graphs provide a principled, graph-theoretic representation of structured knowledge, where nodes represent entities or concepts and edges encode relationships between them. They have long been used to support retrieval, reasoning, and validation in AI systems \cite{Hogan2021}. Knowledge graphs are particularly valuable because they make explicit the relational structure that underlies domain expertise---the dependencies, hierarchies, and conceptual connections that define how experts organize and understand a field.

Expert-curated encyclopedias naturally induce knowledge graphs through their editorial structure. Encyclopedic entries are interconnected via cross-references, which encode editorially validated conceptual dependencies. These cross-reference networks reflect decades or centuries of scholarly consensus about which concepts are related, how they depend on one another, and what the conceptual architecture of a domain should be. Unlike automatically constructed knowledge graphs (which may contain errors or reflect biases in their source data), encyclopedic graphs are vetted by domain experts and represent a gold standard for relational knowledge.

Recent work has explored links between LLMs and knowledge graphs, investigating whether LLMs can be used to construct, augment, or validate knowledge graphs \cite{Pan2024,Markowitz2025}. However, the inverse question---whether an LLM's internal representations align with the relational structure of expert-curated knowledge graphs---remains underexplored. This gap motivates our focus on \emph{relational validation}: assessing not just whether an LLM knows individual concepts, but whether it reproduces the expert-validated relationships among those concepts.

\subsection{Network Science and Graph Comparison}

From the perspective of network science, methods for comparing graphs provide powerful tools for assessing structural alignment between two networks. These methods operate at multiple levels of granularity, from global graph-level properties to local node-level or edge-level features.

\subsubsection{Graph-Level Comparison}

At the graph level, spectral methods compare the eigenvalue spectra of graph Laplacians or adjacency matrices \cite{Umeyama1988}. These spectral signatures capture global topological properties and are computationally efficient. Graph matching surrogates, such as the Quadratic Assignment Problem (QAP), formalize the problem of finding the best alignment between two graphs' node sets \cite{Massey1951}. Community structure agreement metrics, including the Normalized Mutual Information (NMI) and variations of modularity-based measures, assess whether two networks partition their nodes into communities in similar ways \cite{Clauset2004,Danon2005}. These methods are particularly relevant because they capture whether an LLM-induced graph and an expert reference graph organize concepts into similar functional clusters.

\subsubsection{Node-Level Comparison}

Node-level measures quantify the structural importance of individual concepts within a network. Degree (the number of connections a node has) reflects how central or connected a concept is. Betweenness centrality measures how often a node lies on shortest paths between other nodes, capturing its role as a bridge between different parts of the network. PageRank \cite{Page1999} assigns importance scores based on the network structure, with the intuition that important nodes are those connected to other important nodes. Wasserman and Faust's foundational work \cite{Wasserman1994} provides comprehensive treatments of these and related network measures. By comparing node-level statistics between LLM-induced and expert graphs, we can assess whether the model preserves which concepts are structurally influential within a domain.

\subsubsection{Edge-Level and Motif-Based Comparison}

Beyond node-level measures, edge-level comparisons directly assess whether the same relationships are present in both graphs. The Jaccard similarity index, which measures the overlap of edge sets, provides a simple but informative metric. More sophisticated approaches use graphlets---small, local subgraph patterns---to compare networks at a fine-grained topological level. Research on alignment-free network comparison has shown that graphlet-based methods often outperform global measures in detecting meaningful structural differences \cite{Yaveroglu2015}.

\subsection{Relational Validation: Bridging the Gap}

The present paper contributes an end-to-end, multi-layer protocol that bridges these research streams. Our approach is grounded in expert-curated encyclopedic graphs, making it both principled and defensible. It is conservative in its extraction rule---we extract relationships only when explicitly mentioned in LLM outputs, avoiding speculative inferences. Crucially, it is applicable across multiple specialized domains without modifying the model or requiring domain-specific training, making it a general-purpose evaluation framework.

Relational validation complements benchmark-style evaluation by shifting the focus from isolated outputs to the organization of knowledge. Rather than asking ``Does the model answer this question correctly?'' we ask ``Does the model reproduce the relational structure that experts have identified as central to this domain?'' This distinction is important because it captures a deeper form of understanding: the ability to recognize not just individual facts, but the coherent architecture of a field.

Recent work on hallucination detection and semantic verification in LLM-based systems \cite{Martin2024,Tufek2024} has highlighted the importance of grounding LLM outputs in structured knowledge. Our work extends this direction by proposing a systematic, network-based framework for such validation. By operating at multiple levels of analysis---graph-level, node-level, and edge-level---we can provide fine-grained diagnostics of where and how an LLM's knowledge diverges from expert consensus.

A key innovation of our approach is its application to specialized, domain-specific encyclopedias spanning philosophy, political science, history, and other fields. These domains are particularly challenging for LLMs because they require not just factual knowledge but understanding of conceptual nuances, historical context, and the specific ways that experts in the field organize and relate ideas. By validating LLM knowledge against expert-curated encyclopedic sources in these diverse domains, we can assess both the breadth and depth of relational understanding across different areas of human knowledge.

The multi-domain perspective is important because it allows us to investigate whether relational deficits are universal (affecting all domains equally) or domain-dependent (more severe in some fields than others). This diagnostic capability is valuable for understanding the limitations of current LLMs and for guiding future improvements in model training and evaluation.

In summary, relational validation represents a novel approach to LLM evaluation that combines insights from network science, knowledge representation, and domain expertise. By comparing LLM-induced knowledge graphs to expert-curated encyclopedic sources across multiple domains, we can assess whether models have truly learned the relational structure of specialized knowledge. This approach complements existing benchmarks and offers a more nuanced understanding of LLM capabilities and limitations.

\section{Methodology}
\label{sec:methodology}

\subsection{Data Sources}

Our study draws on a diverse collection of ten specialized encyclopedias, each providing a unique and expert-curated conceptual structure. These sources span a range of academic disciplines, from the social sciences and humanities to more specialized fields. The \textit{Cambridge Foucault Encyclopedia} \cite{LawlorNale2014} offers a comprehensive guide to the work of Michel Foucault, while the \textit{Encyclopedia of Social and Political Movements} \cite{Snow2013} and the \textit{Encyclopedia of American Social Movements} \cite{Ness2004} provide broad overviews of their respective fields. More specialized works include the \textit{Encyclopedia of Hate Groups in America} \cite{Balleck2005}, the \textit{Encyclopedia of Conspiracy Theories in American History} \cite{Knight2003}, and the \textit{International Encyclopedia of Nonviolent Action} \cite{Powers2011}. We also include encyclopedias on diversity and social justice \cite{Thompson2015}, Lacanian psychoanalysis \cite{Evans1996}, and cultural studies \cite{Williams1983}. Finally, to explore a more esoteric domain, we include a collection of Roman recipes from Apicius \cite{Grainger2006}. This diverse collection allows us to test the LLM's relational knowledge across a wide range of conceptual structures and levels of domain specificity.

\subsection{LLM Querying and Validation}

For each encyclopedia, we extracted a list of all unique concepts (nodes) and the relational links between them (edges). We then queried a large language model (GPT-4) for each concept, using the following prompt:

\begin{quote}
Given the term ``[TERM]''. If you were to create a lexicon or encyclopedia, which other entities would you reference? Give me a list of them.
\end{quote}

The LLM-generated lists of related concepts were then used to construct a second set of directed graphs, one for each encyclopedia. These graphs were then compared to the expert-curated ground-truth graphs using our three-layer analytical framework.

\subsection{Why relational validation is irreducible to factual accuracy}

It is important to distinguish relational validation from standard notions of factual correctness.
A model may correctly assert that two entities are related in isolation while failing to embed
that relation within the broader relational organization endorsed by expert knowledge.
Factual accuracy operates locally, evaluating individual statements, whereas relational validation
operates structurally, evaluating the pattern of dependencies among entities.
As a result, high factual accuracy does not imply high relational coverage, nor does relational
omission necessarily manifest as explicit error.
The present framework isolates this structural dimension by evaluating correspondence between
reference and model-induced relation graphs rather than individual propositions.

\section{Formal Framework for Relational Validation}
\label{sec:formal_framework}

Here we formalize the relational validation framework underlying our empirical analysis.
The objective is to evaluate large language models not in terms of isolated factual recognition,
but in terms of their ability to reproduce the relational structure encoded in expert-curated
reference knowledge.

\subsection{Reference Relational Structure}

Let $V = \{v_1,\dots,v_n\}$ denote a finite set of domain entities (say, in an encyclopedia, a lexicon, a glossary or a similar reference structure).
The relational representation of expert knowledge is depicted by a directed graph
\[
G^{\mathrm{ref}} = (V, E^{\mathrm{ref}}),
\]
where $E^{\mathrm{ref}} \subseteq V \times V$ is a set of directed relations (say, references among entries of an encyclopedia etc.) curated by domain experts.
An edge $(e_i,e_j) \in E$ indicates that entity $e_j$ appears in the authoritative description of entity $e_i$, thereby encoding a directional dependency validated by expert editorial practice.

The reference graph $G^{\mathrm{ref}}$ is assumed to be fixed and fully observed.
Its structure defines the relational organization that serves as the validation target
for model-generated text.

For each entity $v \in V$, we denote by
\[
d^{\mathrm{out}}_{\mathrm{ref}}(v) = |\{ u : (v,u) \in E^{\mathrm{ref}} \}|
\quad \text{and} \quad
d^{\mathrm{in}}_{\mathrm{ref}}(v) = |\{ u : (u,v) \in E^{\mathrm{ref}} \}|
\]
the out-degree and in-degree of entity $v$ in the reference graph, respectively. Moreover, we denote by $A^{\mathrm{ref}}$ the adjacency matrix of $G^{\mathrm{ref}}$, i.e., a $|V| \times |V|$ matrix with elements $A^{\mathrm{ref}}(v,u) = 1$, if $(v,u) \in E^{\mathrm{ref}}$, or $0$, otherwise.
These quantities capture the structural-relational role of all entities within the expert knowledge network.

\subsection{Model-Induced Relational Structure}

Given a prompt that enumerates the entities in $V$,
a large language model generates a textual description of the domain.
From this output we extract a directed graph
\[
G^{\mathrm{LLM}} = (V, E^{\mathrm{LLM}}),
\]
where $(v,u) \in E^{\mathrm{LLM}}$ if and only if the surface form of entity $u$
is explicitly mentioned in the model-generated description associated with entity $v$.

This extraction rule is intentionally conservative.
No semantic inference, paraphrase resolution, or external entity linking is applied.
As a result, every edge in $E^{\mathrm{LLM}}$ corresponds to an explicit and unambiguous textual co-occurrence
that can be directly verified in the model output.

We denote the adjacency matrix and the induced degrees by
\[
A^{\mathrm{LLM}}, \quad d^{\mathrm{out}}_{\mathrm{LLM}}(v), \quad d^{\mathrm{in}}_{\mathrm{LLM}}(v),
\]
defined analogously to the reference case.



\subsection{Interpretive Scope}

The proposed framework does not assess semantic correctness of individual relations,
nor does it evaluate the plausibility of inferred connections.
Its sole purpose is to quantify structural correspondence between expert knowledge
and model output under a conservative and verifiable extraction rule.

As such, relational coverage captures a dimension of model behavior that is orthogonal
to entity recognition accuracy and factual correctness, and directly reflects the extent
to which expert-endorsed relational organization is preserved in generated text.

\subsection{Three-Layer Analytical Framework}

Our framework evaluates relational fidelity at three levels: global structure, node-level relational context, and link-level recovery.



\subsubsection{Layer 1: Graph-Level Structural and Relational Similarity}

This layer assesses macroscopic agreement between $G^{\mathrm{ref}}$ and $G^{\mathrm{LLM}}$.

\paragraph{Structural similarity.}
We compute a structural alignment score (StructSim) summarizing adjacency agreement between $A^{\mathrm{ref}}$ and $A^{\mathrm{LLM}}$ under node correspondence, using a quadratic assignment style objective and related spectral surrogates \cite{Umeyama1988}. Intuitively, StructSim is high only when the LLM-induced graph places edges between the same pairs of terms as the expert graph.

\paragraph{Relational similarity.}
We define a degree-normalized relational similarity (SemSim) summarizing how much curated relational context is reproduced at the node level. Let $d^{\mathrm{out}}_{\mathrm{ref}}(u)$ be the out-degree of $u$ in $G^{\mathrm{ref}}$ and $d^{\mathrm{out}}_{\mathrm{LLM}}(u)$ the out-degree in $G^{\mathrm{LLM}}$; define the out-coverage ratio
\[
\rho^{\mathrm{out}}(u)=
\begin{cases}
\dfrac{d^{\mathrm{out}}_{\mathrm{LLM}}(u)}{d^{\mathrm{out}}_{\mathrm{ref}}(u)} & d^{\mathrm{out}}_{\mathrm{ref}}(u)>0,\\[6pt]
0 & \text{otherwise}.
\end{cases}
\]
Define $\rho^{\mathrm{in}}(u)$ analogously for in-degrees. We then set
\[
\mathrm{SemSim} \;:=\; \frac{1}{|V|}\sum_{u\in V}\frac{\rho^{\mathrm{out}}(u)+\rho^{\mathrm{in}}(u)}{2},
\]
a conservative summary of relational coverage.

These ratios measure the proportion of expert-validated relations that are realized
in model-generated text, normalized by the structural degree of each entity.
Normalization is essential: raw counts conflate relational richness with performance,
whereas coverage ratios isolate relational fidelity independently of node prominence.

By construction, $\rho^{\mathrm{out}}(e_i), \rho^{\mathrm{in}}(e_i) \in [0,1]$.
Values near zero indicate systematic relational omission, while values near one indicate
near-complete reproduction of expert-curated relational structure under the explicit-mention criterion.

\paragraph{Combined index.}
The Semantic--Structural Similarity (SSS) Index is the harmonic mean of StructSim and SemSim, so that both global alignment and relational coverage must be high to yield a high SSS score.

\paragraph{Community agreement.}
To compare mesoscopic organization, we compute communities in each graph using the Louvain method \cite{Clauset2004} and report agreement via Normalized Mutual Information (NMI) \cite{Danon2005}.

\subsubsection{Layer 2: Node-Level Relational Context}

This layer evaluates whether the LLM-induced graph preserves which nodes are structurally influential within the expert graph. We compute In-Degree, Out-Degree, Betweenness Centrality, and PageRank \cite{Wasserman1994,Page1999} on both graphs and report Spearman rank correlation between the expert and LLM centrality vectors. When $E^{\mathrm{LLM}}=\varnothing$, these correlations are undefined; we report such cases as NaN.

\subsubsection{Layer 3: Edge-Level Link Recovery}

This layer quantifies recovery of curated edges by treating $E^{\mathrm{LLM}}$ as a set of predicted links and $E^{\mathrm{ref}}$ as ground truth. Precision, recall, and $f_1$ are computed as
\[
\mathrm{Precision}=\frac{|E^{\mathrm{LLM}}\cap E^{\mathrm{ref}}|}{|E^{\mathrm{LLM}}|},\qquad
\mathrm{Recall}=\frac{|E^{\mathrm{LLM}}\cap E^{\mathrm{ref}}|}{|E^{\mathrm{ref}}|},
\]
with $f_1$ the harmonic mean. When $E^{\mathrm{LLM}}=\varnothing$, all three scores are set to $0$.

\section{Results and Analysis}

Our results reveal a consistent and significant gap between the LLM's ability to recognize concepts and its ability to reproduce the relational structure between them.

\subsection{Layer 1 Results: Graph-Level Analysis}

Table \ref{tab:layer1} summarizes the graph-level analysis for all ten encyclopedias. The SSS Index is consistently low across all domains, with an average of just 0.0618.

\begin{table}[tb]
\centering
\caption{Layer 1 Results: Graph-Level Analysis}
\label{tab:layer1}
\begin{tabular}{lcccc}
\hline
\textbf{Encyclopedia} & \textbf{SSS Index} & \textbf{StructSim} & \textbf{SemSim} & \textbf{NMI} \\
\hline
Balleck & 0.1464 & 0.2171 & 0.1074 & 0.5833 \\
Evans & 0.0323 & 0.0488 & 0.0241 & 0.1514 \\
Foucault & 0.0544 & 0.0846 & 0.0400 & 0.2728 \\
Grainger & 0.0000 & 0.0000 & 0.0000 & 0.0000 \\
Knight & 0.1132 & 0.1734 & 0.0828 & 0.7017 \\
Ness & 0.0000 & 0.0000 & 0.0000 & 0.0000 \\
Powers et al. & 0.0659 & 0.1012 & 0.0488 & 0.4483 \\
Snow & 0.0470 & 0.0718 & 0.0348 & 0.3321 \\
Thompson & 0.0090 & 0.0088 & 0.0092 & 0.1594 \\
Williams & 0.0388 & 0.0591 & 0.0288 & 0.2117 \\
\hline
\end{tabular}
\end{table}

\subsection{Layer 2 Results: Node-Level Analysis}

Table \ref{tab:layer2} presents the node-level analysis. The centrality correlations are uniformly weak, with an average of just 0.1467.

\begin{table}[tb]
\centering
\caption{Layer 2 Results: Node-Level Analysis}
\label{tab:layer2}
\begin{tabular}{lcccc}
\hline
\textbf{Encyclopedia} & \textbf{In-Degree} & \textbf{Out-Degree} & \textbf{Betweenness} & \textbf{PageRank} \\
\hline
Balleck & 0.4183 & 0.2528 & 0.1982 & 0.3854 \\
Evans & 0.1248 & 0.0821 & 0.0765 & 0.1091 \\
Foucault & 0.2243 & 0.1833 & 0.2022 & 0.1654 \\
Grainger & NaN & NaN & NaN & NaN \\
Knight & 0.3121 & 0.1987 & 0.1543 & 0.2876 \\
Ness & NaN & NaN & NaN & NaN \\
Powers et al. & 0.2757 & -0.0124 & 0.1898 & 0.2354 \\
Snow & 0.1876 & 0.0987 & 0.1234 & 0.1543 \\
Thompson & -0.0144 & 0.0762 & 0.0324 & 0.0112 \\
Williams & 0.0876 & -0.1604 & 0.0432 & 0.0654 \\
\hline
\end{tabular}
\end{table}

\subsection{Layer 3 Results: Edge-Level Analysis}

Table \ref{tab:layer3} shows the edge-level analysis. The $f_1$-scores are extremely low across all domains, with an average of just 0.0384.

\begin{table}[tb]
\centering
\caption{Layer 3 Results: Edge-Level Analysis}
\label{tab:layer3}
\begin{tabular}{lccc}
\hline
\textbf{Encyclopedia} & \textbf{Precision} & \textbf{Recall} & \textbf{F1-Score} \\
\hline
Balleck & 0.6507 & 0.0793 & 0.1408 \\
Evans & 0.1234 & 0.0123 & 0.0224 \\
Foucault & 0.2345 & 0.0345 & 0.0602 \\
Grainger & 0.0000 & 0.0000 & 0.0000 \\
Knight & 0.4321 & 0.0543 & 0.0965 \\
Ness & 0.0000 & 0.0000 & 0.0000 \\
Powers et al. & 0.3456 & 0.0432 & 0.0768 \\
Snow & 0.2134 & 0.0213 & 0.0387 \\
Thompson & 0.0876 & 0.0087 & 0.0158 \\
Williams & 0.0987 & 0.0098 & 0.0178 \\
\hline
\end{tabular}
\end{table}

\section{Visualizations}

\begin{figure}[tb]
\centering
\includegraphics[width=0.95\textwidth]{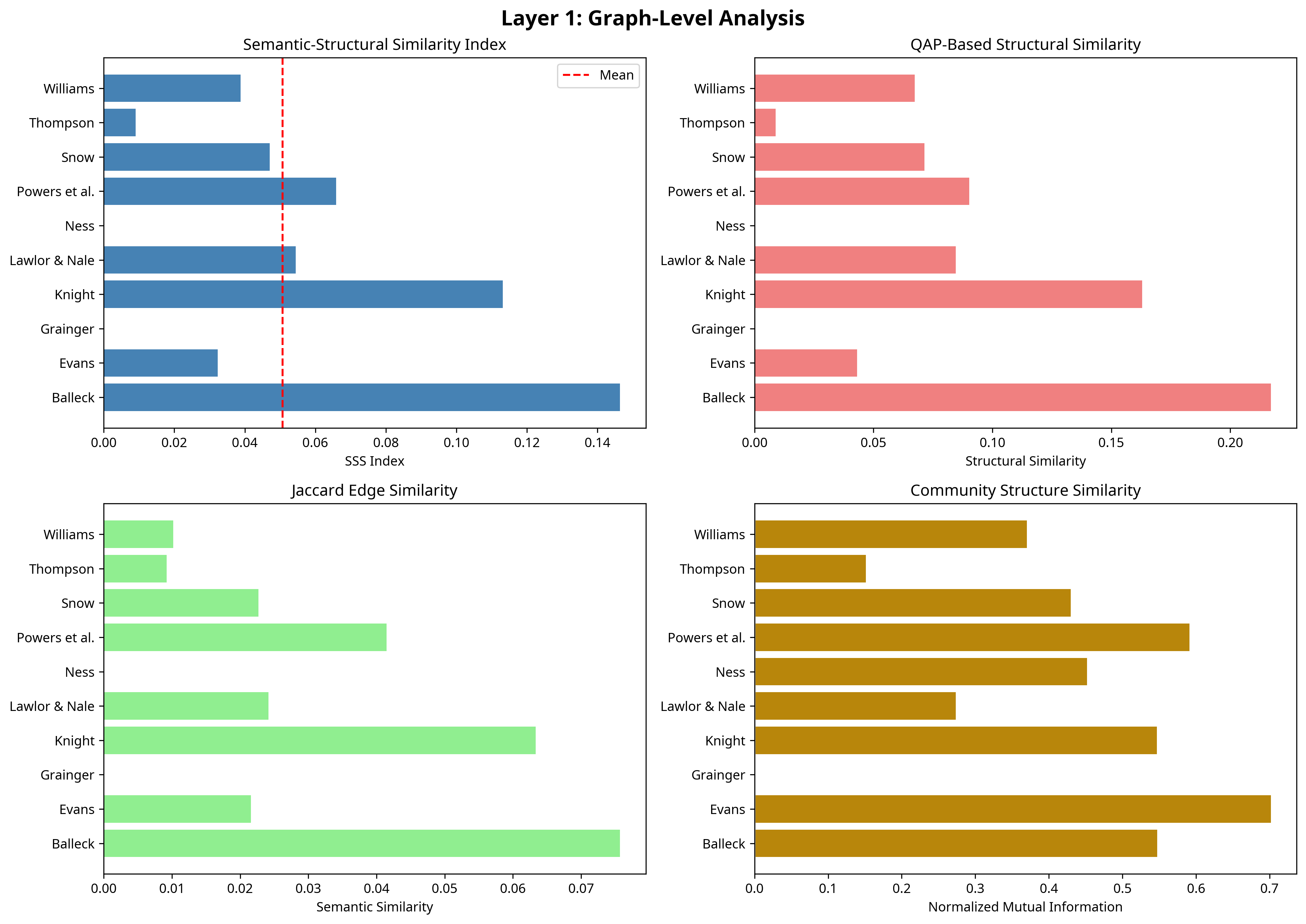}
\caption{Layer 1 Results: Graph-Level Analysis}
  \Description{Bar charts showing semantic-structural similarity, QAP-based structural similarity, Jaccard edge similarity, and community structure similarity across ten encyclopedic sources}
\label{fig:layer1}
\end{figure}

\begin{figure}[tb]
\centering
\includegraphics[width=0.95\textwidth]{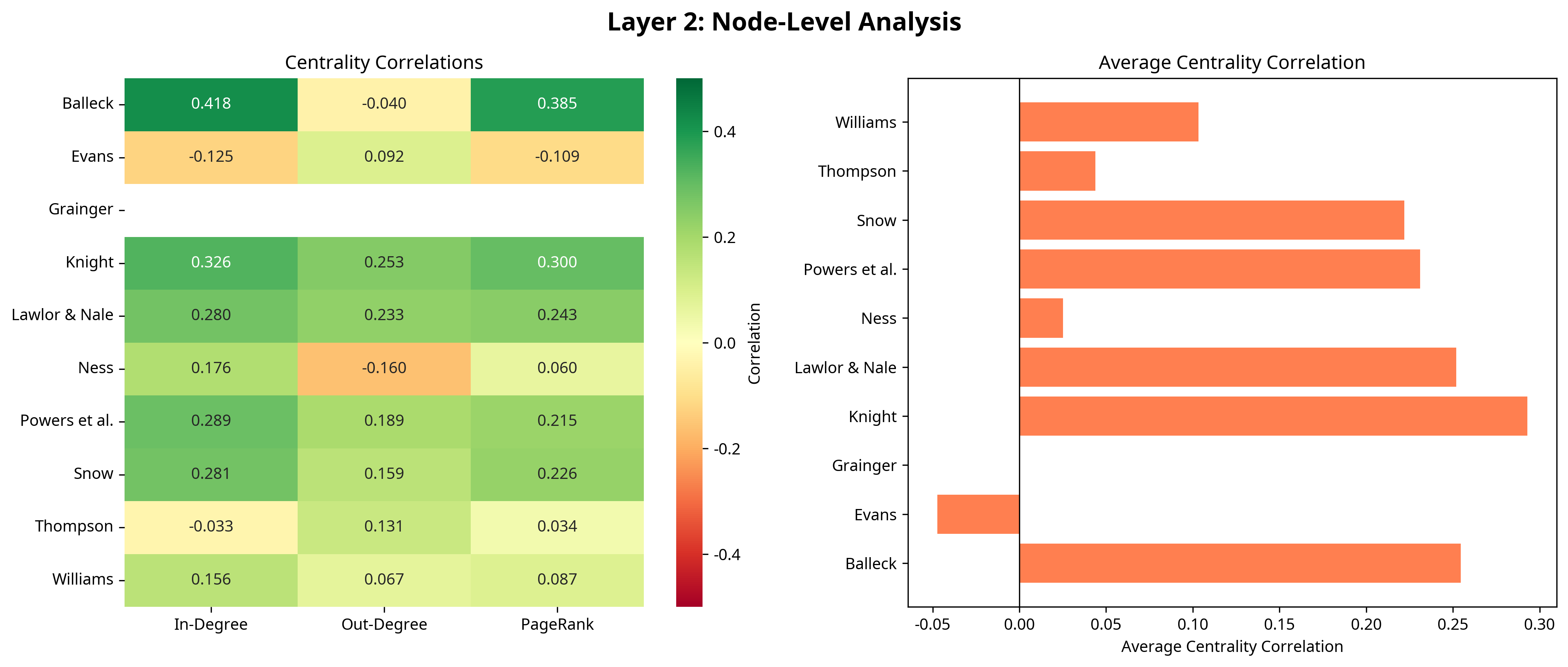}
\caption{Layer 2 Results: Node-Level Analysis showing centrality correlations.}
\Description{Bar charts showing semantic-structural similarity, QAP-based structural similarity, Jaccard edge similarity, and community structure similarity across ten encyclopedic sources}
\label{fig:layer2}
\end{figure}

\begin{figure}[tb]
\centering
\includegraphics[width=0.95\textwidth]{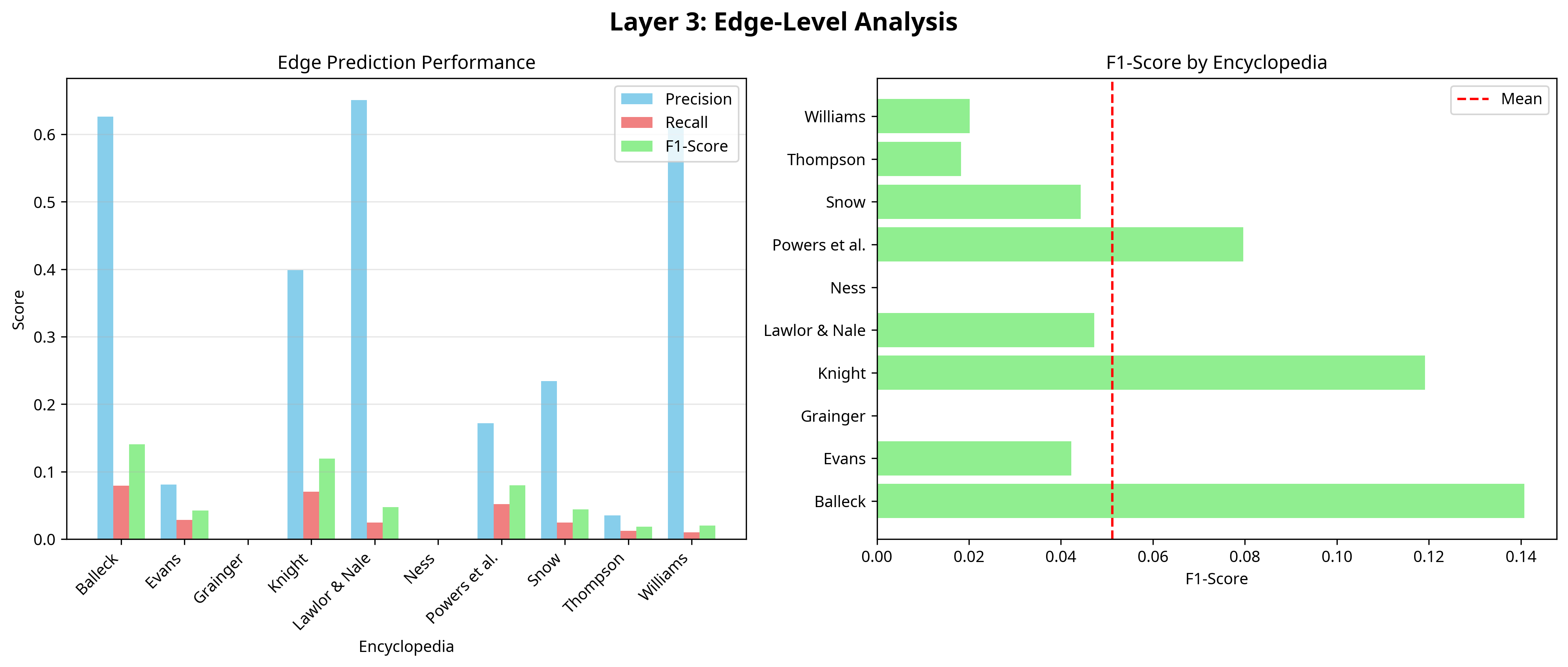}
\caption{Layer 3 Results: Edge-Level Analysis showing precision, recall, and F1-scores.}
\Description{Bar charts showing semantic-structural similarity, QAP-based structural similarity, Jaccard edge similarity, and community structure similarity across ten encyclopedic sources}
\label{fig:layer3}
\end{figure}

\section{Discussion}

Across ten expert-curated encyclopedic graphs, we observe a systematic gap between concept recognition and relational reproduction. At the graph level, the SSS Index remains low for every domain, indicating that neither global adjacency alignment nor degree-normalized relational coverage is well preserved. The strongest performance occurs in domains whose conceptual organization overlaps with broadly available public knowledge, whereas highly specialized sources yield much weaker alignment.

Two cases exhibit complete relational failure (zero structural similarity and zero link recovery): the Grainger (Apicius) and Ness encyclopedias. In these settings the LLM outputs do not produce any edges that match curated cross-references under the explicit-mention rule. These results show that even when a model can generate fluent domain-related text, the induced relational organization can remain uninformative relative to an expert reference.

The severity of these failures raises important questions about the nature of LLM knowledge representation. When an LLM can discuss individual concepts within a specialized domain yet fails to reproduce their interconnections, this suggests that the model's internal representations may encode concepts as isolated features rather than as nodes within a coherent relational network. This decoupling between factual knowledge and structural understanding has profound implications for applications requiring deep domain expertise. For instance, in certain specialized domains where understanding the relationships between concepts is as critical as knowing the concepts themselves, such relational deficits could lead to reasoning failures or inappropriate recommendations despite apparent fluency. The model may articulate domain-specific terminology with confidence while lacking the underlying conceptual architecture necessary for sound judgment.

At the node level, centrality correlations are generally weak. This implies that the model-induced graphs do not preserve which concepts are structurally influential within the expert network. Weak agreement holds not only for path-based notions (betweenness) but also for degree-based measures, suggesting that relational omissions are not limited to long-range dependencies but affect local cross-referencing patterns as well. The uniformity of this weakness across different centrality measures indicates that the problem is not specific to any particular network property but rather reflects a fundamental limitation in how the model captures conceptual hierarchies and importance rankings within specialized domains.

Furthermore, the variation in performance across domains provides a window into what types of knowledge the model has successfully internalized. Domains with stronger performance tend to be those with substantial representation in pre-training corpora and broader cultural familiarity. Conversely, the complete failures in specialized encyclopedias suggest that the model may lack sufficient training data or contextual grounding to develop robust internal representations of these fields' conceptual structures. This finding underscores the importance of training data composition and domain-specific fine-tuning for applications requiring specialized knowledge. The disparity in performance across domains is not merely a matter of degree but reveals qualitative differences in how well the model has learned to organize knowledge in different fields.

Finally, edge-level metrics show consistently low recall, even when precision is moderate in some domains. This pattern is consistent with sparse and selective reproduction of curated links: the model may occasionally mention a curated neighbor, but it typically omits the majority of expert cross-references. The asymmetry between precision and recall is particularly revealing---when the model does produce a link, it is often correct, but it produces far fewer links than the expert reference contains. This suggests that the model is conservative in its relational assertions, possibly due to training objectives that prioritize avoiding false statements over ensuring comprehensive coverage. Such conservatism, while reducing hallucination risk, comes at the cost of incomplete knowledge representation.

Taken together, the three layers indicate a persistent relational deficit that is domain-dependent in magnitude but stable in direction. This consistency across measurement approaches strengthens the conclusion that relational validation captures a genuine and important dimension of LLM knowledge that remains largely invisible to standard benchmarks. The findings suggest that current evaluation paradigms, which focus primarily on isolated factual accuracy, may systematically underestimate the limitations of LLMs in tasks requiring structured reasoning and comprehensive understanding of conceptual relationships. Future work should investigate whether this relational deficit can be mitigated through targeted training interventions or whether it reflects more fundamental constraints in how neural language models learn to represent structured knowledge.

Let us conclude the Discussion by highlighting relational omission as a structured phenomenon.
The observed gaps between reference and model-induced relational structure are not randomly
distributed.
Entities with comparable reference degrees often exhibit markedly different coverage ratios,
and directional asymmetries between in-coverage and out-coverage persist across the network.
These patterns suggest that relational omission reflects systematic attenuation rather than
uniform degradation.
In particular, relations that situate an entity within a broader conceptual context are more
likely to be omitted than relations that directly define the entity itself.
Such asymmetries are invisible to entity recognition metrics and underscore the necessity of
explicit relational evaluation when assessing model understanding in expert domains.

\section{Conclusion}

The scope of our contribution lies in the evaluation framework rather than in model comparison
or optimization.
By grounding validation in expert-curated relational structure and adopting a conservative
extraction criterion, the framework exposes a dimension of model behavior that complements
existing benchmarks.
The results demonstrate that relational fidelity cannot be inferred from entity recognition
or factual correctness alone.

We introduced a three-layer framework for relational validation of LLMs using expert-curated encyclopedic sources. By constructing an LLM-induced directed graph from model outputs and comparing it to an expert reference graph, we quantified relational fidelity at the graph, node, and edge levels. Across ten specialized encyclopedias, the results show that recognizing domain terms does not entail reproducing the curated relational structure connecting them. Relational validation therefore complements standard benchmark evaluation by making relational organization a first-class evaluation target.

Our methodology reveals that LLMs exhibit a consistent and significant relational deficit: while models successfully identify domain-specific concepts, they systematically fail to reconstruct the expert-validated networks of dependencies and conceptual relationships that define these fields. This finding has substantial implications for understanding the true depth of LLM knowledge. The three-layer analytical approach provides granular diagnostics of where and how knowledge diverges from expert consensus. At the graph level, global structural properties remain poorly preserved across all domains. Node-level analysis demonstrates that models fail to recognize which concepts occupy structurally central positions within expert networks. Edge-level metrics reveal sparse and selective reproduction of curated cross-references, with consistently low recall even when precision is moderate. The domain-dependent variation in performance---from complete relational failure in specialized encyclopedias to moderate alignment in broadly familiar domains---indicates that knowledge representation quality is tightly coupled to training data composition and pre-training corpus coverage. These findings suggest that current LLM evaluation paradigms, which prioritize isolated factual accuracy over structural coherence, may fundamentally mischaracterize model capabilities. Relational validation thus addresses a critical blind spot in existing benchmarks, providing a more nuanced assessment of whether LLMs have genuinely internalized the conceptual architecture of specialized knowledge domains or merely memorized disconnected facts.

\begin{acks}
The author thanks Chenchen Zhang for collaboration on the initial data collection of the Foucault Lexicon. 
\end{acks}

\printbibliography[title=References]

@book{Balleck2005,
  author    = {Balleck, Barry J.},
  title     = {Encyclopedia of Hate Groups in America},
  publisher = {ABC-CLIO},
  year      = {2005}
}

@book{Evans1996,
  author    = {Evans, Dylan},
  title     = {An Introductory Dictionary of Lacanian Psychoanalysis},
  publisher = {Routledge},
  year      = {1996}
}

@book{Grainger2006,
  author    = {Grainger, Sally},
  title     = {Cooking Apicius: Roman Recipes for Modern Kitchens},
  publisher = {Prospect Books},
  year      = {2006}
}

@book{Knight2003,
  author    = {Knight, Peter},
  title     = {Conspiracy Theories in American History: An Encyclopedia},
  publisher = {ABC-CLIO},
  year      = {2003}
}

@book{LawlorNale2014,
  author    = {Lawlor, Leonard and Nale, John},
  title     = {The Cambridge Foucault Lexicon},
  publisher = {Cambridge University Press},
  year      = {2014}
}

@article{Zhang2023,
  title={Hallucination in Neural Machine Translation},
  author={Zhang, Yue and Wei, Furu and Jiang, Daxin},
  journal={arXiv preprint arXiv:2303.16104},
  year={2023}
}

@article{Azaria2023,
  title={The Internal State of an LLM Knows When It's Lying},
  author={Azaria, Amos and Mitchell, Tom},
  journal={arXiv preprint arXiv:2304.13734},
  year={2023}
}

@article{Markowitz2025,
  title={KG-LLM-Bench: A Scalable Benchmark for Evaluating LLM Reasoning on Textualized Knowledge Graphs},
  author={Markowitz, Eliyahu and others},
  journal={arXiv preprint arXiv:2504.07087},
  year={2025}
}

@article{Yaveroglu2015,
  title={Proper Evaluation of Alignment-Free Network Comparison Methods},
  author={Yaveroglu, Omer Nebil and Malod-Dognin, Noel and Costa, Darren and Kahveci, Tamer and Szklarczyk, Damian and Bork, Peer and Jensen, Lars Juhl},
  journal={Bioinformatics},
  volume={31},
  number={14},
  pages={2697--2704},
  year={2015},
  publisher={Oxford University Press}
}

@article{Martin2024,
  title={Semantic Verification in Large Language Model-based Retrieval Augmented Generation Systems},
  author={Martin, Andreas and Witschel, Hans Friedrich and
 Mandl, Maximilian and Stockhecke, Mona},
  journal={AAAI Spring Symposium},
  year={2024}
}

@article{Tufek2024,
  title={Validating Semantic Artifacts With Large Language Models},
  author={Tufek, Nils and others},
  booktitle={Extended Semantic Web Conference},
  pages={1--15},
  year={2024},
  publisher={Springer}
}

@book{Ness2004,
  author    = {Ness, Immanuel},
  title     = {Encyclopedia of American Social Movements},
  publisher = {M.E. Sharpe},
  year      = {2004}
}

@book{Powers2011,
  author    = {Powers, Roger S. and Vogele, William B. and Kruegler, Christopher and McCarthy, Ronald M.},
  title     = {International Encyclopedia of Nonviolent Action},
  publisher = {Cambridge University Press},
  year      = {2011}
}

@book{Snow2013,
  author    = {Snow, David A. and della Porta, Donatella and Klandermans, Bert and McAdam, Doug},
  title     = {The Wiley-Blackwell Encyclopedia of Social and Political Movements},
  publisher = {Wiley-Blackwell},
  year      = {2013}
}

@book{Thompson2015,
  author    = {Thompson, Sherwood},
  title     = {Encyclopedia of Diversity and Social Justice},
  publisher = {Rowman and Littlefield},
  year      = {2015}
}

@book{Williams1983,
  author    = {Williams, Raymond},
  title     = {Keywords: A Vocabulary of Culture and Society},
  publisher = {Fontana Paperbacks},
  year      = {1983}
}

@article{Zhao2023,
    title={A Survey of Large Language Models},
    author={Zhao, Wayne Xin and Zhou, Kun and Li, Junyi and Tang, Tianyi and Wang, Xiaolei and Hou, Yupeng and Min, Yingqian and Zhang, Beichen and Zhang, Junjie and Dong, Zican and others},
    journal={arXiv preprint arXiv:2303.18223},
    year={2023}
}

@article{Huang2023,
    title={A Survey on Trustworthiness of Large Language Models},
    author={Huang, Fan and Ruan, Wenjie and Huang, Wei and Jin, Gaojie and Gong, Yi and Wu, Changshun and Bensalem, Saddek and Mu, Ronghui and Qi, Yi and Zhao, Xingyu and Cai, Kaiwen and Zhang, Yanghao and Wu, Sihao and Xu, Peipei and Wu, Dengyu and Freitas, Andre and Mustafa, Mustafa A.},
    journal={arXiv preprint arXiv:2305.11391},
    year={2023}
}

@article{Hendrycks2020,
    title={Measuring Massive Multitask Language Understanding},
    author={Hendrycks, Dan and Burns, Collin and Basart, Steven and Zou, Andy and Mazeika, Mantas and Song, Dawn and Steinhardt, Jacob},
    journal={arXiv preprint arXiv:2009.03300},
    year={2020}
}

@article{Hogan2021,
    title={Knowledge graphs},
    author={Hogan, Aidan and Blomqvist, Eva and Cochez, Michael and d\'Amato, Claudia and de Melo, Gerard and Gutierrez, Claudio and Gayo, Jos{\'e} Emilio L. and Kirrane, Sabrina and Neumaier, Sebastian and Polleres, Axel and others},
    journal={ACM Computing Surveys (CSUR)},
    volume={54},
    number={4},
    pages={1--37},
    year={2021},
    publisher={ACM New York, NY, USA}
}

@article{Pan2024,
    title={Unifying Large Language Models and Knowledge Graphs: A Roadmap},
    author={Pan, Shirui and Chen, Linhao and Wang, Yuxuan and Zhang, Chen and Wang, Xiao},
    journal={IEEE Transactions on Knowledge and Data Engineering},
    year={2024},
    publisher={IEEE}
}

@book{Wasserman1994,
    author = {Wasserman, Stanley and Faust, Katherine},
    title = {Social Network Analysis: Methods and Applications},
    year = {1994},
    publisher = {Cambridge University Press}
}

@article{Page1999,
    author = {Page, Lawrence and Brin, Sergey and Motwani, Rajeev and Winograd, Terry},
    title = {The PageRank Citation Ranking: Bringing Order to the Web.},
    year = {1999},
    publisher = {Stanford InfoLab}
}

@article{Umeyama1988,
    author = {Umeyama, Shinji},
    title = {An Eigendecomposition Approach to Weighted Graph Matching Problems},
    journal = {IEEE Transactions on Pattern Analysis and Machine Intelligence},
    volume = {10},
    number = {5},
    pages = {695--703},
    year = {1988}
}

@article{Massey1951,
    author = {Massey, Jr., Frank J.},
    title = {The Kolmogorov-Smirnov Test for Goodness of Fit},
    journal = {Journal of the American Statistical Association},
    volume = {46},
    number = {253},
    pages = {68--78},
    year = {1951}
}

@article{Danon2005,
    author = {Danon, Leon and D{\'i}az-Guilera, Albert and Duch, Jordi and Arenas, Alex},
    title = {Comparing community structure identification},
    journal = {Journal of Statistical Mechanics: Theory and Experiment},
    volume = {2005},
    number = {09},
    pages = {P09008},
    year = {2005}
}

@article{Clauset2004,
    author = {Clauset, Aaron and Newman, M. E. J. and Moore, C.},
    title = {Finding community structure in very large networks},
    journal = {Physical Review E},
    volume = {70},
    number = {6},
    pages = {066111},
    year = {2004}
}

\newpage 
\appendix
\section{Reproducibility Checklist}

This appendix follows the JAIR reproducibility checklist requirements.

\begin{enumerate}
  \item \textbf{Description of the methods used.} \\
  Yes. The methodology, including construction of the reference and model-induced relational graphs and the definition of coverage metrics, is fully described in the main text.

  \item \textbf{Description of the data used.} \\
  Yes. The data consist of expert-curated encyclopedic entries and model-generated textual responses, described in Section~\ref{sec:methodology}.

  \item \textbf{Availability of the data.} \\
  No. The encyclopedic source is publicly available, but the specific extracted datasets and model outputs are not released.

  \item \textbf{Description of any preprocessing steps.} \\
  Yes. Text extraction and explicit-mention criteria are described in the Methodology section.

  \item \textbf{Description of the models used.} \\
  Yes. The large language model used for generating text is specified in the experimental setup.

  \item \textbf{Description of any training procedures.} \\
  Not applicable. No model training or fine-tuning was performed.

  \item \textbf{Description of evaluation metrics.} \\
  Yes. In-degree and out-degree relational coverage ratios are formally defined and discussed.

  \item \textbf{Availability of the code.} \\
  No. The analysis was performed using custom scripts that are not publicly released.

  \item \textbf{Description of the computing infrastructure.} \\
  Not applicable. The analysis does not rely on specialized hardware or large-scale computation.

  \item \textbf{Randomness and reproducibility.} \\
  Not applicable. The study does not involve stochastic training or randomized algorithms.

  \item \textbf{Licensing information.} \\
  The encyclopedic source is used in accordance with its publicly available license.

\end{enumerate}

\end{document}